\documentclass[aps,prl,twocolumn]{revtex4-1}

\usepackage{graphicx}
\usepackage{relsize}
\def\babar{\mbox{\slshape B\kern-0.1em{\smaller A}\kern-0.1em
    B\kern-0.1em{\smaller A\kern-0.2em R}}}
\def\ie{{\em i.e.}, }
\def\eg{{\em e.g.}, }

\def\msbar{$\overline{\rm MS}$ }

\begin{document}

\title{Precision Constraints on Extra Fermion Generations}

\author{Jens Erler}
\affiliation{Departamento de F\'isica Te\'orica, Instituto de F\'isica, Universidad Nacional Aut\'onoma de M\'exico, 
04510 M\'exico D.F., M\'exico}
              
\author{Paul Langacker}
\affiliation{School of Natural Sciences, Institute for Advanced Study, Einstein Drive, Princeton, NJ 08540, USA}

\date{\today}

\begin{abstract}
There has been renewed interest in the possibility of additional fermion generations. 
At the same time there have been significant changes in the relevant electroweak precision constraints, 
in particular, in the interpretation of several of the low energy experiments.
We summarize the various motivations for extra families 
and analyze them in view of the latest electroweak precision data.
\end{abstract}


\maketitle

In the electroweak (EW) standard model (SM) and most extensions, the number of fermion generations is arbitrary.
It is thus fair to ask whether there may be additional families of quarks and leptons~\cite{Frampton:1999xi}. 
There are interesting theoretical considerations supporting this idea, though most of them
arise in the context of scenarios that hypothesize rather drastic departures from the SM. 

So far, there is no direct experimental evidence either supporting or conflicting with a fourth generation (or anti-generation).
In view of only three observed (nearly) massless neutrinos, however, it is difficult to maintain the notion 
of {\em sequential\/} families of new fermions, although there are examples~\cite{Hill:1990ge} where the appearance 
of a heavy ($m_{\nu'} \gtrsim M_Z/2$) fourth neutrino, $\nu'$, does not appear entirely unnatural.  
On the upside, there is a number of experimental conflicts with the SM expectations at the level of several standard deviations
(too small to be seen as uncontroversial evidence for new physics, yet too large to be ignored)
and some of them could be interpreted as quantum loop effects by the fourth generation states. 

The main point of this letter is the reconsideration of EW precision data in the presence of extra families. 
There are new experimental results from low energy measurements, and there are shifts that occurred due to changes
in the interpretation of previous ones driven in turn by recent progress on the theory side.

One possibility to put a fourth family of quarks, $t'$ and $b'$, to work~\cite{Holdom:1986rn} is 
within models of extended technicolor~\cite{Dimopoulos:1979es}. 
Another one~\cite{Hill:1990ge} is to replace the top quark condensation mechanism~\cite{Bardeen:1989ds} by $t'$ condensation,
since the top is too light for the scenario to work. 
The strongly coupled and condensing fourth generation can also be embedded~\cite{Burdman:2007sx} 
into a warped extra dimension~\cite{Randall:1999ee},
where a heavy $\nu'$ can be arranged for by constructing it as a Dirac fermion 
while the three standard neutrinos are Majorana~\cite{Burdman:2009ih}.

Extra fermions when strongly coupled to the standard Higgs boson may help to generate 
a strongly first-order EW phase transition~\cite{Carena:2004ha} as needed for baryogenesis.
Models of dynamical EW symmetry breaking due to fourth-family quarks and leptons may then also succeed 
in this~\cite{Kikukawa:2009mu}.
The extra quarks could introduce the needed extra CP violation,
which may be enhanced relative to the SM by as much as a factor of $10^{13}$ or more~\cite{Hou:2008xd}.

Finally, string theory vacua typically and easily give rise to even numbers of generations,
while it is usually cumbersome to construct three generation models. 
This has been noted, 
\eg for both free fermionic~\cite{Chaudhuri:1994cd} and orbifold~\cite{Erler:1996zs} string constructions of grand unified theories.

Of course, the Yukawa couplings associated with the new fermions are large.
This may help to achieve non-supersymmetric grand unification~\cite{Hung:1997zj} but may also potentially destabilize 
the Higgs potential or lead to Landau poles below the Planck scale~\cite{Hung:2009hy,Hashimoto:2010at}.

A rough bound on the $t'$ mass, $m_{t'}$, is obtained if one assumes unitarity of the partial S-wave amplitude 
for color-singlet elastic same-helicity $t'\bar{t'}$ scattering already at the tree level~\cite{Chanowitz:1978mv}, 
which for large energies yields~\cite{Marciano:1989ns}
\begin{equation}
\label{unibound}
{m_{t'}^2\over v^2} < {4\pi\over 3}, \quad\quad\quad\quad  m_{t'} < 504 \mbox{ GeV}. 
\end{equation}

The CDF Collaboration set the very recent bound, $m_{b'} > 338 \mbox{ GeV}$~\cite{Aaltonen:2009nr}, from $b' \to t W^\mp$,
complementing and helped by their previous limits, $m_{b'} > 268$~GeV~\cite{Aaltonen:2007je} from $b' \to q Z^0$ and
$m_{q'} > 311$~GeV~\cite{Aaltonen:2007je} from $q' \to q W^\mp$, and bypassing the points raised in Ref.~\cite{Hung:2007ak}.

The CP violating decay rate asymmetry, 
\begin{equation}
\label{akpi}
{\cal A}_{K^\pm \pi^0} \equiv 
\frac{\Gamma(B^- \to K^- \pi^0) - \Gamma(B^+ \to K^+ \pi^0)}{\Gamma(B^- \to K^- \pi^0) + \Gamma(B^+ \to K^+ \pi^0)}, 
\end{equation}
was determined by the \babar, Belle, and CLEO Collaborations to an average of
${\cal A}_{K^\pm \pi^0} = + 0.051 \pm 0.025$~\cite{Amsler:2008zzb}. 
The analogously defined isospin rotated asymmetry, 
${\cal A}_{K^\pm \pi^\mp} = - 0.098 \pm 0.013$~\cite{Amsler:2008zzb}, differs from ${\cal A}_{K^\pm \pi^0}$ by 5.3~$\sigma$,
strongly contradicting the na\"ive expectation ${\cal A}_{K^\pm \pi^0} \approx {\cal A}_{K^\pm \pi^\mp}$.  
The Yukawa matrices for the four family case may be a remedy~\cite{Hou:2005hd} since $Z$ boson penguin diagrams 
and the parameter choice, $m_{t'} \simeq 300$~GeV and $V_{t's}^* V_{t'b} \simeq 0.03\, e^{i 75^\circ}$,
can move ${\cal A}_{K^\pm \pi^0}$ (but not ${\cal A}_{K^\pm \pi^\mp}$) to basically zero, explaining the larger part of the effect.
Based on this, a large time-dependent CP violation in the $B_s^0$ system was predicted~\cite{Hou:2006mx}. 
Subsequently the CDF and D\O\ Collaborations measured this asymmetry in $B_s^0 \to J/\Psi\phi$
and found good agreement with this prediction and with each other, 
but disagreement with the SM, albeit only at the 2.4~$\sigma$ level when the results are combined~\cite{Barberio:2008fa}.
Measurements of other time-dependent CP asymmetries give qualitatively similar results.

Overall the experimental situation is not conclusive and in flux, and so is the optimal parameter choice.
For recent accounts of flavor physics in view of a fourth family, see Refs.~\cite{Soni:2010xh,Buras:2010pi}.
For more details on both the theoretical and experimental situation and for statements about physics beyond the SM with four families,
see Ref.~\cite{Holdom:2009rf}.
                  
The main purpose of this letter is to address the question 
whether the EW data add to the hints that are perhaps implied by the flavor sector.
We employ the oblique parameters, $S$, $T$, and $U$~\cite{Peskin:1990zt}, which parametrize effects of heavy new physics,
\ie $M_{\rm new} \gg M_Z$, contributing to the $W$ and $Z$~self-energies without coupling directly to the ordinary fermions.
For what follows, it is important to recall that new physics models usually come with additional free parameters, 
$N_{\rm par}^{\rm new}$, relative to those in the SM, $N_{\rm par}^{\rm SM}$, 
and this decreases the number of effective degrees of freedom used in a fit, 
$N_{\rm eff} = N_{\rm obs} - N_{\rm par}^{\rm SM} - N_{\rm par}^{\rm new}$, 
where $N_{\rm obs}$ is the number of observables. 

We start our discussion with a case for which $N_{\rm par}^{\rm new} = 0$, 
so the $\chi^2$ minimum, $\chi^2_{\rm min}$, for three and four families can be compared directly.
This occurs when the new quarks and leptons form degenerate doublets 
and corresponds to $S = 2/3\pi = 0.2122$, $T = U = 0$. 
For the Higgs boson mass, $M_H = 112$~GeV (we fix $M_H$ at its 95\% CL lower limit~\cite{Bechtle:2008jh} from LEP~2 whenever
otherwise it would be driven below it), 
we obtain $\chi^2_{\rm min} = 75.54$ compared to $\chi^2_{\rm min} ({\rm SM}) = 43.84$ in the SM ($S = T = U = 0$ by our definition), 
so this case is excluded at the 5.6~$\sigma$ level (we have $N_{\rm eff} = 44$).
Equivalently, one can interpret a fit to $S$ as a fit to the number of degenerate generations and one obtains $N_F = 2.86 \pm 0.20$. 
This agrees with a fit to the number of active neutrinos, $N_\nu = 2.995 \pm 0.007$ (for the same $M_H$) 
when interpreted as the generation number.
One concludes from $N_\nu$ that $m_{\nu'} \gtrsim M_Z/2$, and from the $S$ parameter fit (which is applicable to the heavy $\nu'$ case)
that the good agreement of $N_F$ with the SM value $N_F = 3$ would be coincidental if a fourth family existed.

This restriction can be relaxed drastically by allowing $T$ to vary, since $T > 0$ is predicted by nondegenerate extra doublets. 
Fixing $S = 2/3\pi$, the global fit favors a contribution to $T$ of $0.21 \pm 0.04$ (for $M_H = 112$~GeV) with 
$\chi^2_{\rm min}/N_{\rm eff} = 46.90/43$.
This is due to the strong correlation (87\%) of $S = 0.03 \pm 0.09$ and $T = 0.07 \pm 0.08$.
The central values move to $S = - 0.03~(-0.10)$ and $T = 0.14~(0.29)$ when $M_H$ is increased to 246~(800)~GeV.
Thus generically, the data favor small or negative values of $S$ and $T > 0$.
For example, 
this is the case for nonchiral (vector-like) extra doublets ($S = 0$) which are most consistent with a moderate $T = {\cal O}(0.1)$.
The goodness of the fit, $\chi^2_{\rm min}/N_{\rm eff} = 42.66/43$, is very similar to that of the SM.
If, moreover, the nonchiral matter is also degenerate as predicted in many grand unified theories and other extensions of the SM, 
it does not contribute to any of the oblique parameters and does not require large coupling constants.
Such multiplets may occur in partial families, as in $E_6$ models, or as complete vector-like families~\cite{Martin:2009bg}.

\begin{figure}[b]
 \includegraphics[width=225pt] {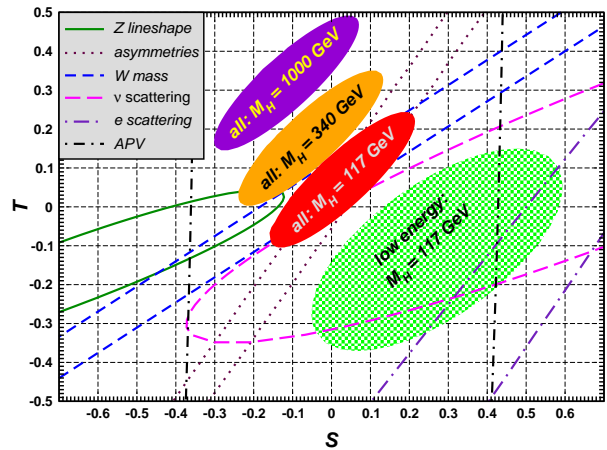}
 \caption{\label{STplot} 
Individual 1~$\sigma$ constraints (39.35\%) on $S$ and $T$.
The contours assume $U = 0$ and $M_H = 117$~GeV except for the central and upper 90\% C.L. filled contours ($\Delta \chi^2 = 4.605$) allowed by all data, which are for the indicated values. 
$\alpha_s$ is additionally constrained by the $\tau$ lifetime.
Since the theory has changed, the strongly $\alpha_s$-dependent solid (dark green) contour 
from $Z$~line shape and cross section measurements~\cite{LEPEWWG:2005ema} 
has moved significantly towards negative $S$ and $T$ compared to our previous analysis~\cite{Amsler:2008zzb}.
The long-dashed (magenta) contour from $\nu$ scattering has moved closer towards the global averages. 
The long-dash-dotted (indigo) contour from polarized $e$ scattering~\cite{Anthony:2005pm,Young:2007zs}
is near the upper tip of an elongated ellipse centered at $(S,T) = (-15,-21)$. 
The dash-dotted (black) contour from APV now agrees perfectly with the SM 
after the completion of a state-of-the-art atomic theory calculation~\cite{Porsev:2009pr}.
The shaded (light green) 1~$\sigma$ ellipse shows the combined low energy data (APV and lepton scattering).}
\end{figure}

But for chiral fermions, $S$ cannot be made that small. 
To elucidate the parameter space we define the 90\% C.L. by the 90\% C.L. allowed region in $(S,T)$ [{\em cf.\/} Fig.~\ref{STplot}], 
and assume in what follows that $m_{\nu'} = 101$~GeV~\cite{Achard:2001qw} and $m_{b'} = 338$~GeV are fixed at their lower limits.
Then we find $S > 0.107$, where the smallest $S$ occurs in a corner of parameter space simultaneously saturating the limits, 
$M_H < 475$~GeV and $T < 0.38$.
In addition, this case has the new charged lepton, $l'$, strongly split from the $\nu'$, $m_{l'} - m_{\nu'} = 140$~GeV, while
we find for the quarks, $m_{t'} - m_{b'} = 28$~GeV.
Our $M_H$ bound is at best only marginally consistent with extra family models which have a strongly interacting Higgs boson
(assuming the absence of other contributions to $S$, $T$, and $U$).
There is a larger allowed parameter space for a light Higgs boson mass,  $M_H = 112$~GeV.
It is bounded by $T < 0.24$ (saturated for $S = 0.19$) and $S < 0.216$ (for $T = 0.218$),
and contains the smallest possible $T = 0.099$ which is reached for $m_{t'} = m_{b'}$ and $m_{l'} - m_{\nu'} = 75$~GeV.
It also contains the best fit which we find for $S = 0.137$, $T = 0.157$, $m_{l'} - m_{\nu'} = 91$~GeV, and 
$m_{t'} - m_{b'} = 14$~GeV. 
Thus the data prefer the leptons to be more split than the quarks, although near the global minimum 
$\chi^2$ is quite shallow along the direction with $m_{l'} + m_{t'}$ approximately constant,
so our splittings are not inconsistent with those found in Ref.~\cite{Kribs:2007nz}. 
The best fit has $\chi^2_{\rm min}/N_{\rm eff} = 43.98/40$,
so that contrary to statements made occasionally in the literature, there is no choice fitting the four family hypothesis better than the SM,
even though $N_{\rm par}^{\rm new} = 4$ parameters have been added.
The important exception is the tuned scenario of a stable $\nu'$ with mass very close to $M_{Z'}/2$~\cite{Maltoni:1999ta,He:2001tp}.
We conclude that a fourth family is disfavored but we find that there is more allowed parameter space 
than with earlier data sets~\cite{Amsler:2008zzb}, which only allowed rather tuned scenarios even at the 90\% C.L.
The reasons can be found mainly in developments in the low energy precision physics,
of which we now briefly discuss the two most important ones.

For decades, measurements of $Z$ induced atomic parity violation (APV) in cesium~\cite{Wood:1997zq} implied $S < 0$,
at times at the 2~$\sigma$ level.  
Several improvements in the atomic theory~\cite{Porsev:2009pr,Ginges:2003qt} 
--- needed to extract the EW physics ---
have now moved $S$ to values well consistent with zero.
In addition, the NuTeV result~\cite{Zeller:2001hh} for $\nu$-nucleus deep inelastic scattering in terms of the on-shell weak mixing angle, 
$s^2_W = 0.2277\pm 0.0016$, was initially 3~$\sigma$ higher than the SM prediction, $s^2_W = 0.22292 \pm 0.00028$.
Since then a number of experimental and theoretical developments shifted the extracted $s^2_W$,
most of them towards the SM:
(i) NuTeV also measured~\cite{Mason:2007zz} a non-vanishing strange quark asymmetry, shifting $s^2_W$ by about $-0.0007$. 
(ii) The measured branching ratio for $K_{e3}$ decays enters 
in the determination of the $\nu_e (\bar\nu_e )$ contamination of the $\nu_\mu (\bar\nu_\mu)$ beam.
Since the time of Ref.~\cite{Zeller:2001hh} it has changed by more than 4~$\sigma$ and the corresponding $s^2_W$ by $+0.0016$.
(iii) Parton density functions seem to violate isospin symmetry much stronger than expected,
implying a shift, $\delta s^2_W = - 0.0026$~\cite{Martin:2004dh,Bentz:2009yy,Gluck:2005xh}.
(iv) The isovector EMC effect~\cite{Cloet:2009qs} reduces $s^2_W$ by $-0.0019$~\cite{Bentz:2009yy}.
With these corrections we find $s^2_W = 0.2242 \pm 0.0018$ (we also increased the error).
The contributions of these and other data sets to $S$ and $T$ are illustrated in Fig.~\ref{STplot}.

 \begin{figure}[t]
 \includegraphics[width=225pt] {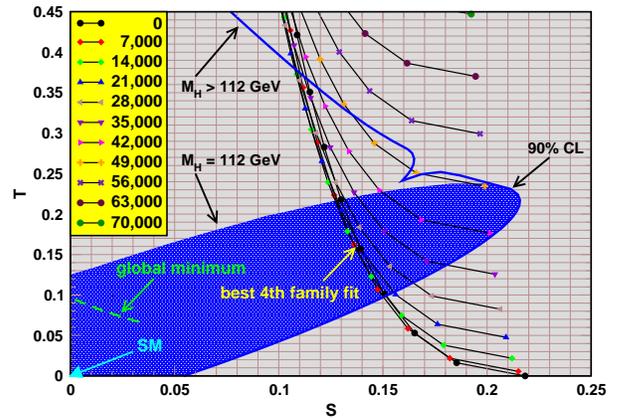}
 \caption{\label{STzoom} 
$S$ and $T$ for various mass splittings.
The 90\% CL ellipse is for $M_H =112$~GeV, while the solid line is the envelope for $M_H \geq 112$~GeV with the kink arising from the
Tevatron exclusion window, 131~GeV $< M_H < 204$~GeV~\cite{Aaltonen:2010sv}, in the presence of a fourth generation.
Each symbol refers to a choice of $m_{t'}^2 - m_{b'}^2$ (with $m_{b'} = 338$~GeV) which is increased in steps of 7,000~GeV$^2$
starting with degeneracy (black circles).
Likewise, moving from right to left increases $m_{\l'}^2 - m_{\nu'}^2$ (with $m_{\nu'} = 101$~GeV)
by the same increments, where the third entries correspond to the choice of Ref.~\cite{Kribs:2007nz}.}
\end{figure}
 
We have assumed $U = 0$, since we have verified that in most of the relevant parameter space $U < 0.03$, 
and where it exceeds this we find $U < 0.11\, T$. 
In any case, allowing $U \neq 0$ decreases $S$ and $T$ (it is negatively correlated with them) which is disfavored.
Similarly, we set the small non-linear oblique parameters, $V$, $W$, and $X$~\cite{Burgess:1993mg}, to zero.
This is currently a sufficiently accurate approximation but we point out 
(i) that exact one-loop results are complete only after their inclusion and their determination from low energy data;
(ii) that the difference between the use of differences and derivatives in the definitions for $S$, $T$, and $U$
is formally of the order of ignoring $V$, $W$, and $X$; and 
(iii) that at that level of precision one should employ \msbar rather than pole quark masses 
which reduces the $T$ parameter by ${\cal O}(10\%)$. 

We were so far considering situations with  $m_{\nu'}$ and $m_{b'}$ at their direct lower bounds.  
One can scale the lepton or quark masses without affecting $S$ and $U$ (in our approximation)
while $T$ scales with the square of the masses. 
This would increase $\chi^2_{\rm min}$ and strengthen our $M_H$ bound, 
but can bring some mass combinations into play [points strictly {\em below\/} the allowed contour in Fig.~\ref{STzoom}].

We also assumed that generation mixing is absent.
A nonzero mixing angle, $\theta_{34}$, between the third and fourth families~\cite{Chanowitz:2009mz} 
give positive and negative definite contributions to $T$ and the $Z \to b\bar{b}$ decay rate, respectively, both worsening the fits.  
The $T$ effect can be eased by allowing larger $M_H$ but at the expense of aggravating the $S$ constraint. 
In fact, we exclude the scenario with $M_H = 810$~GeV and $(S,T) = (0.15,0.48)$~\cite{Chanowitz:2009mz} 
for which we find $\chi^2_{\rm min} = 53.34$ after allowing yet another parameter ($\theta_{34}$). 
We traced most of the disagreement with Ref.~\cite{Chanowitz:2009mz}, where a much milder increase in $\chi^2$ was found, 
in about equal parts to the low energy and more recent high energy data,
an increase in the hadronic vacuum polarization contribution (and decrease in its uncertainty) 
due to more complete and up-to-date experimental and theoretical results~\cite{Davier:2009ag,Davier:2009zi}, and
the implementation of radiative corrections~\cite{Erler:1999ug}. 
Thus, the "three prong composite solution"~\cite{BarShalom:2010bh} with Cabbibo-sized mixing,
and the Higgs boson as well as the $t'$ and $b'$ quarks all close to their unitarity bounds, is strongly conflicting with EW data.
Furthermore, the aforementioned parameters~\cite{Hou:2005hd} to address the asymmetry~(\ref{akpi}) are no longer 
viable~\cite{Chanowitz:2009mz} and have to be adjusted to smaller mixing, and the flavor sector considerations become less convincing
(there are also constraints from flavor changing neutral currents~\cite{Bobrowski:2009ng}).
The remaining parameter space is also difficult to reconcile with gauge coupling unification.

As always, loopholes remain.   
Since the three prong composite solution is really a theory of dynamical symmetry breaking with a composite Higgs sector
(and not just a four-generation extension of the SM), it comes with all the complications of this kind of scenario.
Then the discussion of EW constraints becomes less quantitative for the lack of precise predictions for $S$ and $T$.
A more detailed analysis is required if the $\nu'$ is not a Dirac fermion or only couples to the $\nu_\tau$,
in which cases the L3 $m_{\nu'}$ bounds~\cite{Achard:2001qw} are weaker and slightly negative $S$~\cite{Antipin:2009ks}, 
$T$~\cite{Kniehl:1992ez,Holdom:1996bn}, and $U$~\cite{Kniehl:1992ez} contributions are possible. 

We conclude that while the EW precision constraints have eased somewhat,
a fourth family remains disfavored given that adding up to five new parameters to the SM still deteriorates the global fit.
The part of the parameter space which passes the oblique parameter space at the 90\% C.L. is at odds
with large $M_H$ scenarios as in technicolor-type models. 
It also implies smaller mixing than one would like in face of the flavor physics issues. 
To truly address the latter, we encourage a global EW plus flavor analysis with all sectors, loopholes, and refinements considered and 
with a critical view of how the favored parameter space compares with the expectations from the various motivations discussed earlier.

\begin{acknowledgments}
The work of P.L. is supported by NSF grant PHY--0503584.
J.E. is supported by CONACyT project 82291--F and in part by the German Academic Exchange Service (DAAD).
He greatly acknowledges the hospitality and support extended by the Institute for Theoretical Physics E of the RWTH Aachen 
(where this work was initiated) and discussions with Werner Bernreuther and J\"urgen Rohrwild.
\end{acknowledgments}

\bibliography{4thFamily.bib}

\end{document}